\documentclass[12pt]{article}

\baselineskip
\newcommand{\half}{\frac 1 2 }

\newcommand{\ie}{{\em i.e.} }
\newcommand{\beq}{\begin{equation}}
\newcommand{\eeq}{\end{equation}}
\newcommand{\be}{\begin{eqnarray}}
\newcommand{\ee}{\end{eqnarray}}

\newcommand{\pref}[1]{(\ref{#1})}

\catcode`\@=11
\newcommand{\ccite}[1] {\@ifundefined{b@#1}{\bf ?}{\@nameuse{b@#1}}}
\catcode`\@=12

\begin{document}
\rightline{OSLO-TP 2-99}

\vspace*{2cm}
\centerline{\large \bf SPIN AND STATISTICS}
\vskip 1mm
\centerline{\large \bf FOR QUANTUM HALL QUASI-PARTICLES
%\centerline{\Large\bf Spin and statistics for
%quantum Hall quasi-particles}
\footnote{To appear in the proceedings of the conference
Orbis Scientiae 1998, Fort Lauderdale, \\
December 18 -- 21.}
}
\vskip 15mm
\centerline{\bf Jon Magne Leinaas}
\medskip
\centerline{Department of Physics, University of Oslo}
\centerline{P.O. Box 1048 Blindern, N-0316 Oslo, Norway}

\vskip 15mm
\centerline{\bf ABSTRACT}
In two space dimensions the possibilities of fractional spin as well
as fractional statistics exist. I examine the relation between
fractional spin and statistics for Laughlin quasi-particles in a
two-dimensional electron system with spherical geometry. The
relevance of this for quasi-particles in a planar system is
discussed.
\vspace{.5cm}

\section{Statistics and spin in two dimensions}
I would like to begin by reminding you of the fact that in two space
dimensions there is a richer set of possibilities than in higher
dimensions as far as statistics and spin of particles is concerned.
Quantum statistics is determined by the symmetry of the wave function
under interchange of particle coordinates, and in three and higher
dimensions the corresponding symmetry group is the permutation
group. However, when particle interchange is viewed as a continuous
process under which the coordinates are changed, then the symmetry
group in two dimensions is larger, it is the two-dimensional braid
group rather than the permutation group \cite{Leinaas77}. An element
of this group does not only specify the permutation of the
particles, but also the windings of the particle trajectories under
the interchange of the positions. In dimensions higher than two
these windings can be disentangled, since only interchanges
corresponding to different permutations of the particles are
topologically distinct. This is not possible in two dimensions.

For particles on the plane the coordinates can be written as complex
variables, $z=x+iy$, and for two particles the symmetry under
interchange of the particle positions can be expressed as
\be
\psi\left(e^{in\pi}(z_1-z_2)\right)=e^{in\theta}\psi(z_1-z_2)\,\, ,
\ee
where only the relative coordinate has been written out explicitly.
In this expression $n$ is the winding number of the particle
trajectory in 2-particle space, and $\theta$ is the parameter that
specifies the statistics. The symmetry follows from the assumption
that all configurations which differ only by an interchange of the
particle positions are physically indistinguishable. The wave
function for these configurations should therefore differ at most by
a phase factor. Also for more than two (identical) particles the
symmetry factors have the form $exp(in\theta)$ and they define a
one-dimensional representation of the braid group for the particles.
In two dimensions $\theta$ is a free parameter, while in higher
dimensions it is restricted to the values $\theta = 0 \,\, (mod \,\,
2\pi)$ for bosons and $\theta = \pi \,\,(mod \,\,2\pi)$ for fermions.
For values of $\theta$ different from these two the particles are
said to satisfy intermediate or fractional statistics, and they are
referred to as anyons.

Also spin is different in two dimensions. In three dimensions the
intrinsic spin of a particle is associated with the rotation group
$SO(3)$. It is regarded as the generator of rotations in the rest
frame of the particle. As is well known, the unitary representations
of the rotation group $SO(3)$ restrict the allowed values of the
spin to integer or half-integer multiples of $\hbar$. For particles
in two dimensions the rotation group is reduced to $SO(2)$. This is a
one-parameter group with unitary representations
\be
U(\phi)=e^{i\phi S/\hbar}\,\, ,
\ee
where $\phi$ is the rotation angle. In this case there is no
restriction on $S$, it can take any real value\footnote{I am here
actually referring to representations of the covering group of
$SO(2)$, which are the relevant ones for quantum mechanics.}.

Thus, statistics as well as spin can be regarded as continuous
variables in two dimensions. An obvious question to ask is whether
these two variables are linked by some kind of spin-statistics
relation. This question has previously been discussed in different
ways, and we know from theoretical constructions that many simple
explicit models of two-dimensional particles have such a relation.
Here I will consider this question in connection with a concrete
realization: quasi-particles in the fractional quantum Hall effect.
These quasi-particles are believed, on one hand to be real physical
realizations of anyons in a quasi two-dimensional electron system,
on the other hand to be well described (in some cases) by simple
many-electron wave functions. The question of spin and statistics of
these quasi-particles can therefore be examined rather directly, and
has been done so in the past. One specific study is due to Einarsson
et. al. \cite{Einarsson95}, and my talk is inspired by this paper and
can be seen as a comment to their result.

%%%%%%%%%%%%%%%%%%%%%%%%%%%%%%%%%%%
\section{Spin-statistics relations}
%%%%%%%%%%%%%%%%%%%%%%%%%%%%%%%%%%%
Since we are considering a non-relativistic system, I would like to
stress the point that we cannot expect to find a spin-statistics
theorem that on general grounds gives a strict relation between these
two particle properties. After all we have a simple counter-example
to the standard relation between spin and statistics: spinless
fermions described by one-component anti-symmetric wave functions.
In the context of non-relativistic many-particle theory there seems
to be no problems with such a construction, and this is so for
particles in two as well as in three space dimensions. Nevertheless,
as soon as one leaves the simple point particle description and
makes explicit models where the spin as well as the statistics can
be derived from more fundamental fields, the standard
spin-statistics relation seems naturally to appear in
three-dimensional systems while a linear extension of this relation
appear in two dimensions. Let me just mention some examples from two
dimensions.

A simple electromagnetic model of an anyon is an electric point
charge $e$ with an attached magnetic flux $\phi$, that is confined to
a small region around the charge. (The mechanism that binds the flux
to the charge is not so important and neither is the detailed
profile of the magnetic field surrounding the charge.) In addition
to the Coulomb interaction between such charge-flux composites,
there will be an Aharonov-Bohm interaction between the charge of one
composite and the flux of the other. When two  composites are
interchanged the latter gives rise to a phase factor that can be
identified with the statistics factor. A simple calculation gives
for the statistics parameter
\be
\theta=-\frac{e\phi}{\hbar c}\,\, .
\ee
There is an electromagnetic spin associated with a charge flux
composite, due to the overlap of the electric and magnetic fields.
Using the expression for electromagnetic angular momentum reduced to
its two-dimensional form, we calculate the spin to be
\be
S=- {1\over c} \int d^2r B \vec r \cdot \vec E = -\frac{e\phi}{2\pi
c}\,\, .
\ee
We note that the statistics parameter and the spin both are
determined by the same quantity $e\phi$.

A second example is provided by soliton solutions in the $O(3)$
non-linear $\sigma$-model with a topological (Hopf) term
\cite{Wilczek83}. In this case the strength of the topological term
determines the spin as well as the statistics of the solitons. A
third example is given by the particles described by a scalar field
theory with Chern-Simons coupling \cite{Hansson88}. The Chern-Simons
field gives an explicit realization of fractional statistics in the
form of an Aharonov-Bohm effect. It also affects the conserved
angular momentum and thereby links the spin to the statistics of the
particles.

In the examples referred to above (as well as in some other examples)
the relation between spin and statistics has the simple form
\be
S=\left[\frac{\theta}{2 \pi} \,\, (mod \,\,1) \right] \,\hbar\,\, .
\label{ss1}
\ee
It coincides with the standard relation for bosons ($\theta=0$) and
fermions ($\theta =\pi/2$) and extends that linearly to all other values
of the statistics parameter $\theta$.

Even if the simple relation \pref{ss1} is favoured by many anyon
models, we do not have a clear specification of the general
conditions under which the relation should be satisfied. There do
exist, however, some general arguments for a less restrictive form
of the spin-statistics relation that are based on the assumption
that there exist both anyons and anti-anyons in the system under
consideration. Let me briefly give the arguments for this
generalized spin-statistics relation, since it is relevant for the
quantum Hall quasi-particles.

We then assume that there exist fractional statistics particles of a
type we denote by $p$ (with some unspecified statistics parameter
$\theta$). There also exist another type of particles $\bar p$, that
we consider as anti-particles to $p$. Since we are not considering a
relativistic theory, we do not assume charge conjugation symmetry
(symmetry between $p$ and $\bar p$). The important point is the
assumption that a $p-\bar p$ pair can be created and annihilated
inside the system. This means that all long range effects of a single
particle are canceled by the corresponding effects of an
anti-particle. This has consequences for statistics as well as for
spin.

For a $p-\bar p$ pair there are no long-range Aharonov-Bohm effects.
That means that the phase factor introduced by transport of another
particle of type $p$ around the pair is the trivial factor $1$ for a
path far away from the two particles. If these two particles also are
sufficiently far apart, the phase factor can be written as a product
of one factor from each of the particles in the pair. We write this
as
\be
exp\left(2i\,(\theta_{pp}+\theta_{p \bar p})\right)=1 \,\, .
\label{cond1}
\ee
We easily see that $\theta_{pp}$ is identical to the statistics phase
$\theta$ of particles $p$. The other phase $\theta_{p \bar p}$ is
sometimes referred to as a mutual statistics phase. It describes an
Aharonov-Bohm interaction between two non-identical particles $p$ and
$\bar p$. Clearly we have a similar condition when a particle of type
$\bar p$ is transported around the pair,
\be
exp\left(2i\,(\theta_{\bar p \bar p}+\theta_{\bar p p})\right)=1 \,\, .
\label{cond2}
\ee
The two conditions \pref{cond1} and \pref{cond2}, and the symmetry
relation $\theta_{\bar p p}=\theta_{p \bar p}$, mean that all
phases can be expressed in terms of a single phase $\theta$,
\be
\theta_{\bar p \bar p}=\theta_{pp}&=&\theta\,\,\,\, (mod\,\, \pi)
\nonumber
\\
\theta_{\bar p p}=\theta_{p \bar p}&=&-\theta\,\,\,\, (mod\,\,\pi)
\,\, .
\ee

A rotation of the $p-\bar p$ pair by an angle $2 \pi$ also has to
give rise to a trivial phase factor. We write this as
\be
exp\left(2\pi \frac{i}{\hbar}(L_{cm}+L_{rel}+S_p+S_{\bar p})\right)=1
\,\, .
\label{rot}
\ee
The orbital angular momentum has here been divided into a
center-of-mass part $L_{cm}$ and a part determined by the relative
motion, $L_{rel}$; $S_p$ and $S_{\bar p}$ are the intrinsic spins of
the two particles. $L_{cm}$ has integer eigenvalues in multiples of
$\hbar$, while the spectrum of $L_{rel}$ is shifted due to the
nontrivial phase $\theta_{p \bar p}$. The eigenvalues are
$(n-\theta/\pi)\,\hbar,n=0,\pm 1,\pm 2...$. With this inserted in
\pref{rot} we get
\be
\half (S_p+S_{\bar
p})=\left[\frac{\theta}{2\pi}\,\,(mod\,\,\half)\,\right]
\hbar \,\, .
\ee
This is the generalized spin statistics relation. It only involves
the sum of the spins of the anyon and the anti-anyon. Even if these
two spins are equal we note the relation is less restrictive than
the relation \pref{ss1}. It does not exclude spinless fermions or
bosons with half-integer spin.

%%%%%%%%%%%%%%%%%%%%%%%%%%%%%%%%%%%%%%%%%%%%
\section{Anyons in the quantum Hall system}
%%%%%%%%%%%%%%%%%%%%%%%%%%%%%%%%%%%%%%%%%%%%
The quasi-particles of the quantum Hall system are charged
excitations in a 2-dimensional electron gas subject to a strong
perpendicular magnetic field. In general the quasi-particles are
fractionally charged and obey fractional statistics; they are
charged anyons in a strong magnetic field. For special filling
fractions of the lowest Landau level, $\nu=1/m$, $m$ odd, there exist
simple (trial) wave functions, originally introduced by Laughlin
\cite{Laughlin83}, for the ground state of the many-electron system
as well as for the quasi-particle excitations. Expressed in complex
electron coordinates, the (non-normalized) $N$-electron ground state
has the form
\be
\psi_m(z_1,z_2,...,z_N)=\prod\limits_{i<j} {(z_i-z_j)^m}e^{-{1 \over
{4\ell^2}}\sum\limits_{k=1}^N {\left| {z_k} \right|^2}} \,\, ,
\label{gs}
\ee
with $\ell=1/\sqrt{\frac{\hbar c}{eB}}$ as the magnetic length, and
$eB$ taken to be positive. The one quasi-hole state is
\be
\psi_{Z}^{qh}(z_1,z_2,...,z_N)=\prod\limits_{i=1}^N
{(z_i-Z)}\psi_m(z_1,z_2,...,z_N) \,\, ,
\label{qh}
\ee
with $Z$ as the position of the quasi-hole.  Multi-hole wave
functions are constructed in a similar way, with several prefactors
of the form given in Eq.\pref{qh}. For the oppositely charged
quasi-electron Laughlin has suggested a wavefunction of the form
\be
\psi_{Z}^{qe}(z_1,z_2,...,z_N)=\prod\limits_{i=1}^N
{(\frac{\partial}{\partial z_i}-Z^*)}\psi_m(z_1,z_2,...,z_N) \,\, .
\label{qe}
\ee
Supported by general arguments, as well as numerical studies, the
ground state and the quasi-hole state are believed to be very well
represented by the wave functions \pref{gs} and \pref{qh} (in a
homogeneous system). However there is an asymmetry between the
quasi-hole and the quasi-electron, and one should note that there is
not a similar strong evidence in favour for the quasi-electron wave
function
\pref{qe}\footnote{For a recent discussion see
Ref.~\cite{Kjonsberg99}.}.

The form of the quasi-particle wave functions determine the
fractional charge as well as their fractional statistics. This was
demonstrated by Arovas, Schrieffer and Wilczek who calculated the
Berry phases associated with shifts of the quasi-particle
coordinates along closed curves \cite{Arovas84}. Let me give a brief
comment on this in general terms.

The wave functions for configurations with $M$ quasi-holes define a
$M$ (complex) dimensional submanifold in the $N$-electron Hilbert
space parameterized by the quasi-hole coordinates. A fractional
statistics representation (or anyon representation)
\cite{Laughlin90} of the system can be introduced in terms of wave
functions defined on this manifold, $\psi(Z_1,Z_2,...,Z_M)$. The
$M$-dimensional manifold, on which the wave-functions are defined
can be interpreted as the configuration space (alternatively as the
phase space) of the (classical) $M$ quasi-hole system. In a
low-energy approximation we may consider the system restricted to
this space. The kinematics as well as the dynamics of the quasi-hole
system are determined from the $N$-electron system by projection on
the complex submanifold. In particular, the kinematics is determined
from the geometry of the manifold, and the charge and the statistics
appear as geometrically determined parameters.

The scalar product of the $N$-electron Hilbert space defines, by
projection, a complex geometry in the $M$-dimensional quasi-hole
space. It is expressed in terms of the Hermitian matrix
\be
\eta_{kl}=\langle D_k \psi|D_l\psi\rangle \,\, ,
\ee
with
\be
D_k=\partial_k+iA_k,\,\,\,A_k=i\langle\psi|\partial_k
\psi\rangle \,\, .
\ee
$|\psi\rangle$ denotes the $M$-quasi-hole state and $\partial_k$ is
the partial derivative with respect to a set of real coordinates in
the quasi-hole space. $A_k$ is the Berry connection defined by the
set of quasi-particle states. The real (and symmetric) part of
$\eta_{kl}$ determines a metric on the $M$ quasi-particle space
\be
g_{kl}=Re\langle D_k \psi|D_l\psi\rangle \, \, ,
\ee
while the imaginary (and anti-symmetric) part determines a symplectic
form, that we identify as the ``Berry magnetic field",
\be
b_{kl}=2Im\langle \partial_k \psi|\partial_l\psi\rangle =\partial_k
A_l-\partial_l A_k \, \, .
\ee

For a single quasi-hole the form of $\eta_{kl}$ is strongly
restricted by translational and rotational invariance (in the limit
$N\rightarrow \infty$) and by analyticity in the variable $Z$ ,
\be
\eta_{kl}=-\frac{b_1}{2} (\delta_{kl}+i \epsilon_{kl})\, .
\label{Bmf}
\ee
Here $b_1$ is a constant that can be expressed in terms of the the
real magnetic field, $b_1=\frac{e^* B}{\hbar c}$, with the
coefficient $e^*$ as the effective charge of the quasi-hole. A Berry
phase calculation for a loop in the plane determines the flux of
$b_1$ through this loop, and comparison with the real magnetic flux
then gives the effective charge
$e^*$ \cite{Arovas84}.

For a two quasi-hole state an expression similar to \pref{Bmf} is
valid for $\eta_{kl}$, if this now refers to the relative coordinate
of the two quasi-holes. In this case $b_1$ is replaced by a function
$b_2(R)$ that depends on the relative distance $R$. For small
$R$ the form of this function is determined by local properties of
the quasi-holes. For large $R$,  $b_2(R)$ is expected to approach
rapidly the constant $\half b_1$ when the quasi-holes are well
localized objects. The flux of $b_2$ then has the form
\be
\int_{r<R}d^2r\, b_2(R)=\half \,\pi R^2 \,b_1 - 2\theta \,\, ,
\ee
where $\theta$ is identified as the statistics parameter of the
quasi-holes. Again this parameter can be determined by a Berry phase
calculation, that measures the flux of $b_2(R)$ within a given
radius.

Berry phase calculations based on the quasi-hole wave function
\pref{qh} gives $e^*=-e/m$ for the charge and $\theta=-\pi/m$ for the
statistics parameter, with $e$ as the electron charge
\cite{Arovas84}. For the quasi-electron wave function \pref{qe} one
cannot derive the results so easily \cite{Kjonsberg96}, but the
expected results for the physical quasi-electron is
$e^*=e/m$ and $\theta=2-\pi/m$, as determined from general reasoning
and numerical studies \cite{Canright94}.

Whereas charge and statistics can be determined geometrically, in
terms of Berry phases associated with closed curves of one and two
quasi-particles, the spin cannot be determined quite as easily.
However, as pointed out by Einarsson \cite{Einarsson91} and Li
\cite{Li92} there is a way to derive spin from Berry phases,
provided the particles move in a curved space. If the spin can be
viewed as a three-dimensional spin constrained to point in the
direction orthogonal to the two-dimensional surface, there will be a
contribution to the Berry phase when transporting the quasi-particle
around a loop that is proportional to the product of the spin value
and the solid angle traced out by the spin \cite{Berry84}. This
suggests the following form of the Berry magnetic field
\be
b_1=\frac{e^*B}{\hbar c}-\frac{S}{\hbar}\kappa\,\, ,
\label{BS}
\ee
with $\kappa$ as the Gauss curvature and the coefficient $S$ as the
spin. It is not obvious that calculations of Berry phases for
quasi-holes will give a separation in two terms of this form, but if
they do, the spin can be determined from the Berry phases. This is
the assumption made in \cite{Einarsson95}. In this case a quantum
Hall system with the geometry of a sphere is considered. One should
note that in this case the magnetic field $B$ as well as the
curvature $\kappa$ are constants. That means that there is no clear
distinction between the two contributions to the Berry phase in
Eq.\pref{BS}. However if the charge $e^*$ of the quasi-particle on
the sphere is the same as the quasi-particle charge on the plane
(which seems reasonable), then the second term can be separated from
the first one and the spin can be determined.

%%%%%%%%%%%%%%%%%%%%%%%%%%%%%%%%%%%%%%%%%%%
\section{Quantum Hall states on the sphere}
%%%%%%%%%%%%%%%%%%%%%%%%%%%%%%%%%%%%%%%%%%%
In practice, to create a quantum Hall system with the geometry of a
sphere can hardly be done. A radially directed magnetic field is then
needed, and this means that a magnetic monopole should be found and
placed at the center of the sphere. However as a theoretical
construction a spherical Hall system can easily be created, and as
first shown by Haldane such a geometry may conveniently be used in
the study of certain aspects of the quantum Hall effect
\cite{Haldane83}. Also for numerical calculations it is convenient
due to the lack of boundaries
\cite{Canright94}.

To have a consistent quantum description of the electrons in the
monopole field, Dirac's quantization condition has to be satisfied,
\be
\frac{e\phi}{4\pi \hbar c}=\half N_{\phi} \,\, ,
\ee
where $\phi$ is the total flux of the monopole field and $N_{\phi}$
is an integer. This means that the total magnetic flux through the
sphere is quantized in units of the flux quantum
$\phi_0=\frac{hc}{e}$,
\be
\phi= N_{\phi}\; \phi_0 \,\, ,
\ee
with $N_{\phi}$ as the number of flux quanta.

Laughlin states like \pref{gs},\pref{qh} and \pref{qe} can be
constructed on the sphere and can conveniently be expressed in terms
of the coordinates $u=\cos(\theta/2)$
and $v=\sin(\theta/2)exp(i\phi)$, with $\theta$ and $\phi$ as the
polar coordinates on the sphere. The form of the ground state is (in
the Dirac gauge $e\vec A=eB\,\tan\frac{\theta}{2}\,\,
\vec e_{\phi}$)
\be
\psi_m= \prod_{i<j}(u_i v_j - u_j v_i)^m \,\,
,\;\;\;N_{\phi}=m(N-1)\,\, ,
\label{gss}
\ee
and this is non-degenerate, with all particles in the lowest Landau
level, provided the number of electrons $N$ is linked to the number
of flux quanta $N_{\phi}$ as indicated above. If one flux quantum is
added, a hole state is created,
\be
\psi^{qh}_{UV}=\prod_i (V u_i-Uv_i) \psi_m\,\, , \;\;\;
N_{\phi}=m(N-1)+1
\,\, ,
\label{qhs}
\ee
with $(U,V)$ as the quasi-hole coordinates, and if one flux quantum
is removed, a quasi-electron state is created,
\be
\psi^{qe}_{UV}=\prod_i (V^* \frac{\partial}{\partial
u_i}-U^*\frac{\partial}{\partial v_i}) \psi_m\,\, ,\;\;\;
N_{\phi}=m(N-1)-1 \,\, ,
\label{qes}
\ee
now with $(U,V)$ as the quasi-electron coordinates.

For the quasi-hole state a detailed calculation of the Berry phase
has been performed in Ref.~\cite{Einarsson95}, with a discussion of
the different contributions. I will not repeat that here, let me
rather show how the result concerning the spin can be derived
directly from rotational invariance, without reference to Berry
phases. This derivation is based on the assumption that the
quasi-particle can be represented as a particle with charge $e^*$ in
the monopole field.

For a single electron moving in a magnetic monopole field, the
conserved angular momentum has the form
\be
\vec J=\vec r\times{\vec \pi}+\mu \vec {\hat r} \,\, ,
\ee
with $\vec \pi$ as the mechanical momentum,
\be
\vec \pi= \vec p-\frac{e}{c}\vec A \,\, ,
\ee
and
\be
\mu=-\frac{e\phi}{4 \pi c}
\ee
as the component of the total angular momentum in the radial
direction $\vec {\hat r}$. This spin can be identified as the
electromagnetic angular momentum due to the overlap of the electric
field of the charge with the magnetic monopole field. This radially
directed spin is quantized due to the Dirac condition,
\be
\mu=\half N_{\phi}\,\hbar \,\, ,
\ee
and this quantization condition can alternatively be derived directly
from the requirement of rotational invariance, \ie from the condition
that the operator $\vec J$ should generate unitary representations of
the rotation group.

Thus, there are two invariants associated with the angular momentum,
\be
\vec J^2=j(j+1)\hbar^2\,\, , \,\,\, \hat{\vec r} \cdot \vec J =\mu
\,\, ,
\ee
with the restriction
\be
j=|\mu|,|\mu|+1, ...\,\, .
\ee
The smallest value of $j$ can be identified as corresponding to the
lowest Landau level, and as on the plane, the mechanical
part of the angular momentum then has its smallest value.
For $N$ electrons the total angular momentum is the sum of the
contributions from each electron,
\be
\vec J = \sum_{i=1}^N \vec J_i\,\, .
\ee
The ground state \pref{gss} is rotationally symmetric, with $j=0$,
while the spin of the quasi-hole state \pref{qhs} is $j=N/2$.

In the anyon representation the quasi-hole is represented as a
(single) charged particle in the monopole field. If we assume that
it can be treated as a point particle, the angular momentum has the
same form as for a single electron,
\be
\vec J=\vec r\times{\vec \pi}+(\mu^*+S) \, \vec {\hat r} \,\, .
\ee
In this expression $\vec r$ is the quasi-hole coordinate and
$\mu^*=-\frac{e^*\phi}{4\pi c}$ is the radially directed
electromagnetic spin. $S$ is a possible additional radially directed
spin, an intrinsic spin of the quasi-hole. We note that such an
additional spin in fact has to be added in order to preserve
rotational invariance. If $e^*$ is taken to be identical to the
charge $e/m$ of a quasi-hole in a planar system, then
$\mu^*=N_{\phi}/2m$. This is in general not a half-integer, and the
condition for rotational invariance is therefore not satisfied with
$S=0$. The value of $S$ can be determined if we identify the anyon
coordinates with the coordinates $(U,V)$ of the quasi-hole state
\pref{qhs}. The spin component of this state in the $(U,V)$ direction
is $N/2$, and this gives the relation
\be
\frac{1}{2m}N_{\phi}+S=\half N \,\, .
\ee
 With the number of flux quanta related to the electron number as
indicated in Eq.\pref{qhs} this gives the spin value
\be
S_{qh}=\half-{1\over{2m}}=\half+{\theta\over{2\pi}} \,\, .
\label{spinqh}
\ee
where $qh$ now labels the spin of the quasi-hole. This result for the
spin is the same as the one determined by Berry phase calculations
\cite{Li92,Einarsson95}. We note that the spin-statistics relation
given by \pref{spinqh} is not identical to the relation \pref{ss1}
indicated by the anyon models referred to at an earlier stage. There
is an additional term $1/2$ that looks like a shift between the
boson and fermion value of
$\theta$. However, one should also note that the contribution from
the intrinsic spin of the electrons has not been included here. For
fully polarized electrons in the plane this contribution is $-1/2m$.
For large electron numbers, this contribution is presumably the same
on the sphere. Thus, with all contributions included we get
$S_{qh}=\half-{1\over{m}}=\half+{\theta\over{\pi}}$, and we still do
not recover the relation \pref{ss1}. The only exception is for
$m=1$, the case of a fully occupied lowest Landau level. The spin is
then $-1/2$, in accordance with the standard spin-statistics
relation.

The quasi-electron state \pref{qes} can be examined in a similar
way. The spin component in the radial direction in this case has the
opposite sign and there is also a change in the relation between the
number of flux quanta and the electron number. The spin value now is
\be
S_{qe}=\half-{1\over{2m}}=-\half+{\theta\over{2\pi}} \,\, .
\label{spinqe}
\ee
The contribution from the intrinsic spin of the electrons in this
case is $1/2m$, which gives the total spin $S_{qe}=\half$. Also here
the original spin-statistics relation is not satisfied.

However, the two expressions \pref{spinqh} and \pref{spinqe} show
that the generalized spin-statistics relation is satisfied in the
form
\be
\half(S_{qh}+S_{qe})={\theta\over{2\pi}} \,\, .
\ee
That is the case also when the contribution from the intrinsic spin
of the electrons are included, since the contribution to the
quasi-electron spin is the same, but with opposite sign as the
contribution to the quasi-hole spin.

%%%%%%%%%%%%%%%%%%%%%%%%%%%%%%%%%%%%%%%%%%%%%%%%%
\section{Spin on the sphere -- spin on the plane}
%%%%%%%%%%%%%%%%%%%%%%%%%%%%%%%%%%%%%%%%%%%%%%%%%
The spin values \pref{spinqh} and \pref{spinqe} are determined for
quasi-particles on a sphere. What conclusion can we now draw
concerning quasi-particles in a planar system?  Is there a local spin
associated with the quasi-particles with value identical to the one
found on a sphere? The discussion we find in Ref.~\cite{Einarsson95},
and also the results found in a paper by Sondhi and Kivelson
\cite{Sondhi92}, do not support this conclusion\footnote{Somewhat
surprisingly this is not seen as a problem in \cite{Einarsson95},
with the explanation that the spin in the planar system does not
have a dynamical significance.}. Thus, if their conclusions are
correct, there is no simple relation between the spin of the
quasi-particle on the sphere and a spin derived from the angular
momentum of the electrons in a planar system. This is somewhat
disappointing since the main motivation for putting the
quasi-particles on the sphere, I assume, was to be able to visualize
the quasi-particle spin, not to create the spin. The usual picture
of the quasi-particle excitations is that they are strongly
localized in space and that they have particle like properties with
sharply defined quantum numbers such as charge, mass and possibly
spin. If the quasi-particle spin determined on the sphere is not the
same as the quasi-particle spin on the plane, that presumably means
that it cannot be thought of as a {\em local} spin associated with
the quasi-particle. The spin could in principle be due to a small
renormalization of the charge of the quasi-particle when put on a
sphere,
\be
e^*_{sphere}=e^*\left(1+\frac{m-1}{mN}\right) \,\, ,
\ee
However, the $N$ dependence of the correction term does not seem to
fit the picture of the quasi-particle as a strongly localized object.

Let me briefly discuss the question of the quasi-particle spin for a
planar system. The normal component of the conserved angular
momentum of an electron in a homogeneous magnetic field is
\be
J=(\vec r \times \vec \pi)_z +\frac{eB}{2} r^2 \,\, ,
\ee
with $\vec r$ as a vector in the $(x,y)$-plane. The first term is the
mechanical angular momentum of the circulating electron, whereas the
second term can be interpreted as the electromagnetic spin (with an
infinite $\vec r$-independent term subtracted).

For electrons in the lowest Landau level, the conserved angular
momentum can be written in the form
\be
J= \hbar \left( -\int d^2r \rho(r) + \frac{1}{2\ell^2}\int d^2r r^2
\rho(r) \right) \,\, ,
\label{angmom}
\ee
with $\rho$ as the particle density. The first term, the mechanical
angular momentum is proportional to the particle number, since all
electrons in the lowest Landau level carry one unit of (mechanical)
angular momentum. The second term is the contribution from the
electromagnetic angular momentum. It has the opposite sign of the
first term and dominates this so that for all angular momentum
eigenvalues the spin is non-negative.

The total angular momentum \pref{angmom} diverges with the size of
the system, the first term as the electron number $N$ and the second
term as $N^2$. This is so for the ground state \pref{gs} as well as
for the quasi-particle states \pref{qh} and \pref{qe}. Clearly, if a
local, finite spin should be associated with the quasi-particle, one
has in some way to subtract the angular momentum of the ground state.
A simple definition of the quasi-particle spin would be
\be
S_{qp}=\lim_{R\rightarrow \infty}(J_{qp}(R)-J_0(R)) \,\, .
\label{sqp}
\ee
where $J_{qp}(R)$ is the total angular momentum of the quasi-particle
state within a radius $R$ and $J_0(R)$ the angular momentum of the
ground state within the same radius. The size of the electron system
is here regarded as infinite. Even if these two terms diverge
separately for large $R$, the difference should stay finite and give
a well-defined value for the spin.

The first term of the angular momentum \pref{angmom} gives a
contribution to the quasi-particle spin (after the subtraction of the
ground state spin) which is determined by the charge of the
quasi-particle. The contribution is $\pm 1/m$ with + for the
quasi-hole and - for the quasi-electron. The second term is not so
easy to determine as the first term, but in the paper by Sondhi and
Kivelson \cite{Sondhi92} (where a similar definition of the
quasi-particle spin is used), there is a discussion of the
quasi-hole case. In this case the plasma analogy, introduced by
Laughlin, can be applied. In the plasma analogy the square modulus
of the quasi-hole wave function
\pref{qh} is interpreted as the partition function of a Coulomb
system consisting of $N$ free (unit) charges in a homogeneous
neutralizing background, with the presence of an additional fixed
charge of value $1/m$ (the quasi-hole). The integrated particle
number is then determined as the screening charge of this fixed
charge, with the value $-1/m$. Also the second moment of the particle
number density, which is relevant for the second term of the angular
momentum, can be related to the value of the charge. In fact,
assuming that the conditions for ``perfect screening" to be satisfied
\cite{Laughlin90}, there is a cancellation between the two terms of
the the angular momentum so that the quasi-hole spin, as defined
above, vanishes. This is the conclusion of Sondhi and
Kivelson\footnote{Sondhi and Kivelson also consider corrections to
the spin due to the electromagnetic self-interaction of the
quasi-hole. Such corrections are important in order to give the
correct value of the spin for the physical quasi-hole, but have not
been taken into consideration here.}. With this conclusion it it
difficult to see any connection between the physical spin of the
quasi-hole state in the plane and the spin determined on the sphere.
If the physical spin vanishes for any value of $m$ this in fact
rules out any connection between the (physical) spin and the
statistics parameter of the quasi-particles.

However, as a final point I would like to pose the question whether
the conclusion concerning the spin, which is based on the use of the
plasma analogy, is necessarily true, or whether another conclusion
may be possible. Clearly, for a full Landau level, with $m=1$, the
quasi-hole spin vanishes since the hole is created simply by removing
an electron in a spin $0$ state. For $m=3$ the situation is not quite
as obvious and one has to refer to the situation in a one-component
plasma with a $1/3$ charge screened by a plasma of integer charges. I
am not able to judge the claim that the perfect screening condition
is satisfied in this case, but I have noted with interest that in
Ref.~\cite{Baus80} one refers to a ``basic belief" in the underlying
assumption when the perfect screening sum rule is derived.

There is of course a way to avoid the reference to the plasma
analogy. That is to make a straight forward calculation of the spin
\pref{sqp} of the planar system, and I will cite some preliminary
results for Monte-Carlo calculations performed by Heidi Kj{\o}nsberg
for an electron system consisting of $N=100$ electrons. The
numerical calculations reproduce values for the integrated
quasi-hole spin $S_{qh}$, within a variable radius
$R$ around the quasi-hole, which is placed at the center of the
circular electron system defined by the Laughlin wave function.

Let me first give some values for the spin evaluated on the sphere,
as given by Eq. \pref{spinqh}. For
$m=1$ the spin is $0$, for $m=3$ the spin is $1/3$ and for $m=5$ the
spin is $2/5$, all spins expressed in units of $\hbar$.

The numerical results for the the planar system agree well with the
value $0$ for the $m=1$ state. However, for $m=3$ this is not the
case. For values of the radius $R$ that lie between the size of the
quasi-hole and the size of the full electron system, the results
indicate instead a fairly stable value close to $1/3$, that agrees
with the value found on the sphere. For $m=5$ the results are not so
clear, due to larger finite size effects and also due to larger
statistical fluctuations in the Monte Carlo calculations.
Nevertheless, also here the results indicate a spin value different
from $0$ and possibly consistent with the value $2/5$.

So I would like to finish by referring to the question of the spin
of the quasi-hole as an interesting one which deserves a further
study. I feel that the situation in a sense would be more satisfying
if the spin evaluated on the sphere could be identified as the
physical spin of the quasi-particle also for a planar quantum Hall
system. But such a conclusion would raise some new and interesting
questions concerning the use of the plasma sum rules for the
Laughlin states.

%%%%%%%%%%%%%%%%%%%%%%%%%%
\section*{Acknowledgments}
%%%%%%%%%%%%%%%%%%%%%%%%%%
I appreciate the help of Heidi Kj{\o}nsberg who have performed the
numerical calculations referred to in this paper. I am grateful to
Hans Hansson and Anders Karlhede for several helpful comments and
would like to thank their group at the Department of Physics,
Stockholm University, for hospitality during a stay in February 1999.

\end{document}